\documentclass [12pt,twoside]{article}           
\usepackage[left,modulo]{lineno}
\usepackage{setspace}

\countdef\arxiv=201

\ifnum\arxiv=0 \input cpreb.tex \fi

\ifnum\arxiv=1

\usepackage{epsfig,times,lscape}
\usepackage[usenames]{color}

    \ifnum\count102=1

    \fi

    \ifnum\count102=2
\fi

\newcommand{\nc}{\newcommand}

     \ifnum\count101=1
\nc{\qI}[1]{\section{{#1}}}
\nc{\qA}[1]{\subsection{{#1}}}
\nc{\qun}[1]{\subsubsection{{#1}}}
\nc{\qa}[1]{\paragraph{{#1}}}

\def\qpar{\vskip 2mm plus 0.2mm minus 0.2mm}
\def\qL{\hfill \break}
     \fi 

      \ifnum\count101=2
 \nc{\qI}[1]{\parindent=0mm \vskip 8mm 
{\centerline{\LARGE \color{red}#1}}\vskip 3mm}
\nc{\qA}[1]{\vskip 2.5mm \noindent 
{{\bf\large\color{blue}  #1}} \vskip 1mm \parindent=0mm}
 \nc{\qun}[1]{\vskip 1mm \noindent {\sl #1 }\quad }

\def\qL{\hfill \break}
\def\qpar{\vskip 2mm plus 0.2mm minus 0.2mm}

      \fi

\def\qth{\vrule height 12pt depth 0pt width 0pt}
\def\qtb{\vrule height 0pt depth 5pt width 0pt}

\nc{\qfoot}[1]{\footnote{{#1}}}

      \ifnum\count101=1
\def\qbu{\hfill \par \hskip 6mm $ \bullet $ \hskip 2mm}
\def\qee#1{\hfill \par \hskip 6mm (#1) \hskip 2 mm}
      \fi
      \ifnum\count101=2
\def\qbu{\hfill \par \hskip 4mm $ \bullet $ \hskip 2mm}
\def\qee#1{\hfill \par \hskip 4mm (#1) \hskip 2 mm}
      \fi

\def\qparr{ \vskip 1.0mm plus 0.2mm minus 0.2mm \hangindent=10mm
\hangafter=1}

     \ifnum\count101=1 
 \def\qdec#1{\parindent=0mm\par {\leftskip=2cm {#1} \par}}
     \fi
     \ifnum\count101=2
  \def\qdec#1{\parindent=0mm \par {\leftskip=1cm {#1} \par}}
  
  \def\qcitb#1{\noindent \hbox to 102mm{\hfill \small #1} \vskip 1mm}
      \fi

 \def\qpages#1{\count102=0{\loop\advance\count102 by 1
 \null \vfill\eject \ifnum\count102<#1 \repeat}}

\def\qn#1{\eqno \hbox{(#1)}}

\def\qth{\vrule height 12pt depth 0pt width 0pt}
\def\qtb{\vrule height 0pt depth 5pt width 0pt}

\def\qv{\vskip 0.1mm plus 0.05mm minus 0.05mm}
\def\qhu{\hskip 0.6mm}
\def\qhv{\hskip 3mm}

\def\qhw{\hskip 1.5mm}
\def\qleg#1#2#3{\noindent {\bf \small #1\qhw}{\small #2\qhw}{\it \small #3}\qv }
\fi

\begin{document}
\thispagestyle{empty}



\markboth{{\sl \hfill  \hfill \protect\phantom{3}}}
        {{\protect\phantom{3}\sl \hfill  \hfill}}

\color{yellow} 
\hrule height 10mm depth 10mm width 170mm 
\color{black}

\vskip -12mm 

\centerline{\bf \Large Effect of population density on epidemics}

\vskip 10mm

\centerline{Ruiqi Li$ ^1 $,
Peter Richmond$ ^2 $ and Bertrand M. Roehner$ ^3 $}

\vskip 10mm
\large

%
{\bf \color{blue} Abstract}\qL
Investigations of a possible connection between 
population density and the propagation and magnitude
of epidemics have so far led to mixed and unconvincing
results. There are three reasons for that. 
(i) Previous studies did not focus on the appropriate
density interval.
(ii) For
the density to be a meaningful variable the population
must be distributed as uniformly as possible. If an area
has towns and cities where a majority of the 
population is concentrated its average density is
meaningless.
(iii) In the propagation of an epidemic the initial proportion
of susceptibles (that is to say persons who have not
developed an immunity) is an essential, yet usually unknown, factor.
The assumption that most of the population is susceptible
holds only for {\it new} strain of diseases.
\qL
It will be shown that when these requirements are taken care of,
the size of epidemics is indeed closely connected with
the population density. This empirical observation
comes as a welcome confirmation of the classical KMK
(Kermack-McKendrick 1927) model. 
Indeed, one of its key predictions is that the size
of the epidemic increases strongly (and in a non linear way)
with the initial density of susceptibles. 
\qL
An interesting consequence is that, contrary to common beliefs, 
in sparsely populated territories, 
like Alaska, Australia or the west coast of the United states
the size of epidemics among native populations must have been limited
by the low density even for diseases 
for which the natives had no immunity (i.e., were susceptibles).

\vskip 5mm
\centerline{\it \small Provisional. Comments are welcome.}
\centerline{\it \small Version of 10 March 2018}
\vskip 4mm

{\small Key-words: epidemic, propagation, population density,
Kermack and McKendrick model.}

\vskip 5mm

{\normalsize
1: School of Systems Science, Beijing Normal University,
Beijing, China. \qL
Email: liruiqi@mail.bnu.edu.cn  \qL 
2: School of Physics, Trinity College Dublin, Ireland.
Email: peter\_richmond@ymail.com \qL
3: Institute for Theoretical and High Energy Physics (LPTHE),
University Pierre and Marie Curie, Sorbonne Universit\'e,
Centre National de la Recherche Scientifique (CNRS),  
Paris, France. 
Email: roehner@lpthe.jussieu.fr
}

\vfill\eject

\qI{Introduction}

To begin with, let us say that although
the data that we analyze in this paper are mostly
from the early 20th century our objective is {\it not}
to write a historical paper. What guided us
is the fact that the
phenomenon that we wish to study, namely the 
density effect in the propagation of epidemics,
can only be studied on highly infectious diseases.
As one knows, with the possible exception of influenza,
such diseases have been practically eliminated
in developed countries. It is true that they
still exist
in developing countries but in most of these countries the
accuracy of vital statistics is not very good. 
\qpar

In historical accounts of the contacts between native
populations and white immigrants (for instance
in the Pacific ocean islands, in Australia or in the United
States) it is often said that natives were wiped out
by diseases that their immune system could not
fight. Most often such claims are made without
any supporting evidence being presented in terms 
of number of deaths due to specific epidemics. 
As most often native people
were living in sparsely populated areas, the question
of how population density affects the propagation
of epidemics is obviously of central importance.
It is this question which was at the origin of the
present study. We will come back to it in our
conclusion.
\qpar

Seen from the side of the pathogens, contagion is 
a form of diffusion in which the virus or bacteria
jump from one individual to another. If the
transmission takes place through air or water
both intuition and mathematical modeling would suggest
that it is facilitated by a higher population density%
\qfoot{As a second step, at a more detailed level, one 
would of course expect that proximity due to specific human
mobility and interactions will also play a role
(Li et al. 2017a, 2017b).}%
.
The paradox is that in most studies that we know about,
the impact of density cannot be seen clearly.
This is illustrated below by the results of two
papers and by a number of personal observations.

\qA{Papers on influenza epidemics}

In a study of the pandemic of 1918 in England and Wales 
(Chowell et al. 2008) the authors observe
``we did not find any obvious association between death
rates and measures of population density''.
\qpar

Similarly in a study of the same 1918 epidemic in New Zealand
(Haidari et al. 2006) the authors present a plot for
($ x= $ population density, $ y= $ death rate, $ n=108 $) and find
the two variables to be basically uncorrelated ($ r=0.17 $).

\qA{Observations for several contagious diseases}

To these sources we can add the following tests which similarly lead
to mixed and even puzzling results. 

%
\begin{figure}[htb]
\centerline{\psfig{width=17cm,figure=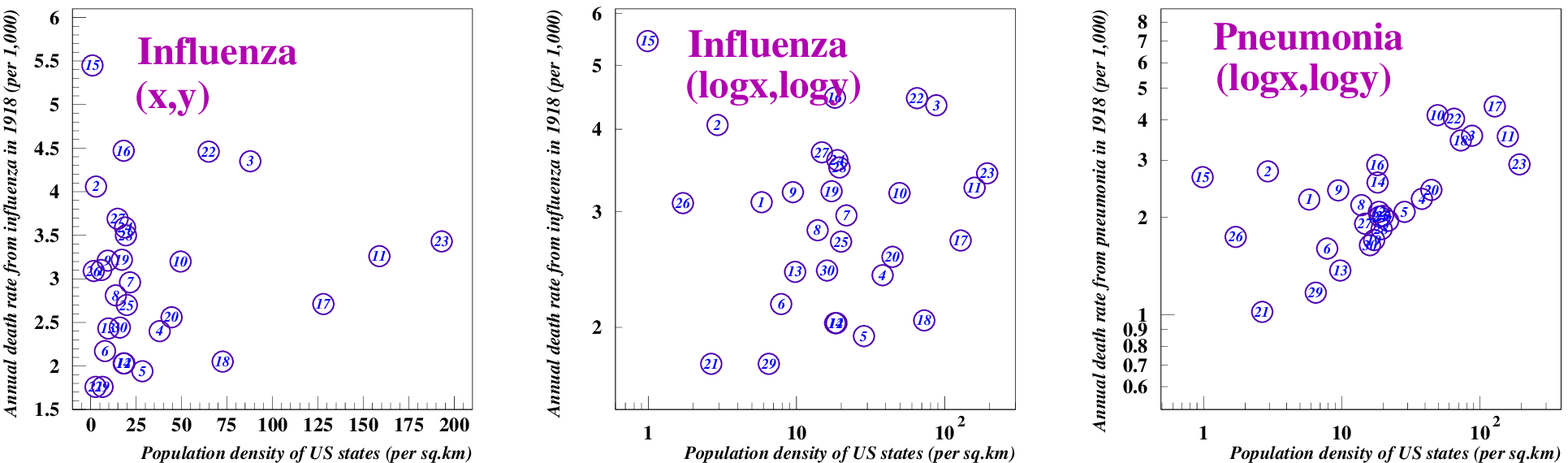}}
\qleg{Fig.\qhu 1a,b,c\qhv Relationship between population density by
state and death rate, USA, 1918.}
{{\bf (a)} 
This graph is for influenza.
There is basically no correlation 
(the correlation is 0.10 and the confidence interval 
is $ (-0.27,0.45) $) which means that no
regression line can be drawn. However, it seems (by comparison
with the pneumonia case) that there 
are some obvious outliers such as: 15=Montana, 2=Colorado,
16=New Hampshire, 22=Pennsylvania, 3=Connecticut. It is
not easy to understand why these states have death rates
that are abnormally high.
{\bf (b)} 
The graph of (a) was redrawn with log scales
The correlation, namely $ -0.068 $ is still not significant.
{\bf (c)} 
This graph is for pneumonia. The correlation ($ \log d, \log\mu $) is
$ 0.62 $, CI$ =(0.33,0.80) $.}
{Sources: Density: Historical Statistics of the United States,
p. 24; death rate: Mortality Statistics 1918, p.118.}
\end{figure}

 The 1918 volume of ``Mortality Statistics''
published by the US Bureau of the Census gives the death rate
from influenza for each of the 30 Registration states
i.e., the states which recorded death statistics.
The (density, death rate) correlation turns out to be equal
to 0.10 which, for a probability level of 0.95, is not
significant in the standard sense that the confidence
interval, namely $ (-0.27,0.45) $ contains 0.
\qpar

As a matter of fact, the scatter plot has the same
shape as the one mentioned above
for New Zealand: for densities under 25 
per square-kilometer 
there is a very large dispersion of death rates;
then for densities over 50 the plot becomes more orderly, yet
with some outliers.
\qpar

In Fig. 1b the broad range of $ d $ suggests to use a log scale.
In this case, however, for the sake of
consistency one must also use a log scale for $ \mu $
in spite of the fact that $ \mu $ has only a fairly narrow range.
The reason is that for $ d \rightarrow 0 $ it is natural
to expect $ \mu \rightarrow 0 $; with two log scales
both $ d $ and $ \mu $ will tend to $ -\infty $. On the contrary,
with $ (\log d, \mu) $ the two variables would have different limits.
\qpar

Do these tests mean that there is no correlation whatsoever between
density and death rate? Not necessarily. It simply means that
the background noise over-rides any weak association that may exist.
   
\qA{Contagious versus non contagious diseases}

Table 1 compares the death rates in large cities with
those in rural areas. Here again the results are
found to be fairly puzzling.
For instance, for contagious diseases, one would
expect the death rate ratio cities/rural to be larger than 1. 
Not only is this ratio just barely higher
than 1 but in addition the ratio for non-contagious
diseases is markedly higher than 1.
The most intriguing result is the one for pneumonia.
Whereas Fig. 1c for 1918 showed a clear excess mortality
in places of high density the results for 1940
(the only year for which such data are given in Linder
et al. 1947) show higher death rates in rural places. 
In addition, if one draws the graph of death rates by states
one finds that the correlation which existed in 1918 has
disappeared in 1940. So, although we ignore the reason
of this change, at least the two results are consistent
with each other. 

%
\begin{table}[htb]

\small
\centerline{\bf Table 1: Comparison of death rates in
cities of more than 100,000 and in rural areas, USA, 1940}

\vskip 5mm
\hrule
\vskip 0.7mm
\hrule
\vskip 0.5mm
$$ \matrix{
&\hbox{Tubercu-}\hfill & \hbox{Pneu-}\hfill & \hbox{Syphilis}\hfill &
\hbox{Average} & \hbox{Intra}\hfill & \hbox{Disease}\hfill &
\hbox{Disease}\hfill & \hbox{Average}\cr
&\hbox{losis}\hfill & \hbox{monia}\hfill & \hbox{}\hfill &
\hbox{(contagious} & \hbox{cranial}\hfill & \hbox{of the}\hfill &
\hbox{of the}\hfill & \hbox{(non contagious}\cr
\qtb
&\hbox{}\hfill & \hbox{}\hfill & \hbox{}\hfill &
\hbox{diseases)} & \hbox{lesion}\hfill & \hbox{heart}\hfill &
\hbox{coronary}\hfill & \hbox{diseases)}\cr
\noalign{\hrule}
\qth
\hbox{Cities}\hfill & 40.8 & 55.5 &11.1 &  & 78.4 & 23.6 & 45.4 & \cr 
\hbox{Rural} \hfill & 34.0 & 70.0 & 8.80 & & 88.0 & 18.6 & 23.0 & \cr
\qtb
\hbox{Cities/Rural} \hfill & 1.20 & 0.79 & 1.26 & 1.08 & 0.89 & 1.27 &
1.97&1.36\cr
\noalign{\hrule}
} $$
\vskip 1.5mm
Notes: The death rates are per 100,000 population.
There is no clear difference between cities and rural areas.
The most surprising result is probably the one for 
pneumonia which, contrary
to expectation, is notably higher in rural places
(may be related to better medical treatment available
in cities).
As a preliminary explanation one may posit
that the lower rural death rate for diseases of coronary arteries 
is due to the fact that life in rural places involves more
physical activity.  
\qL
Source: Linder et al. (1947).
\vskip 2mm
\hrule
\vskip 0.7mm
\hrule
\end{table}
%
\qA{Components of the background noise}

What is the meaning intended for the expression ``background
noise''. An illustration from particle physics may be helpful.
There are currently experiments under way in order to
observe if protons can disintegrate into lighter particles. 
Such an event can be identified
by detecting the particles that it produces. However,
in spite of the fact that in such experiments
the tank is located deep under
ground it is nevertheless hit by particles emitted by
the Sun or by the surrounding rocks. This is what
physicists call background noise. It is different
from purely statistical noise. Whereas the later cannot
be reduced (except by taking averages over large numbers of events),
the background noise can be reduced for instance by 
shielding the tank in appropriate ways. 
In short, the background noise is produced by specific sources
which, once clearly identified, may be eliminated.
\qpar

What are here the factors which contribute to the background noise
for epidemics?\qL
One can mention the following. 
\qee{1} In principle it would be better to consider incidence
rates rather than death rates. By considering death rates
one mixes two effects: the diffusion of the disease and
the availability (and effectiveness) of medical treatment.
For instance death rates from tuberculosis may be higher 
in poor districts where pulmonary
diseases are widespread and where no treatment is provided.
However, death rates may be a good proxy for
incidence rates for sufficiently large areas.
\qee{2} The existence of large cities in an area makes
the average density a fairly biased variable.
\qee{3} The initial percentage of susceptibles which
depends on the previous occurrences of the disease.
\qee{4} The climate, whether hot or cold, dry or humid.
As an illustration of how the climate effect can generate
spurious data it can be mentioned that in the late 19th
and early 20th centuries the dry and sunny climate
of Arizona, Colorado, Nevada and New Mexico attracted
many tuberculosis patients and led to the building
of health facilities (sanatoriums, boarding houses
and even canvas camps). Naturally, this resulted in
highly inflated death rates in the corresponding
states. 
\qee{5} The age structure
of the population. As the 1918 epidemic hit particularly
middle-aged persons, if this group is over-represented 
the total death rate will be higher.
\qpar

The most important lesson to retain for the following sections
is that one should consider {\it large} density changes so that
their impact can overcome the background noise.
As a matter of fact, Chowell et al. (2008) and Haidiri et al. (2006)
also made the observation that urban areas have higher death
rates than rural areas but they did not 
discuss the noise versus signal levels
nor did they specify
what must be done to make the signal stand out more clearly.

\qI{Empirical evidence}

\qA{Overview for contagious diseases}

Population density ($ d $) is a variable with a broad range of variation,
from a few persons per square kilometer in rural areas to
a few thousands in big cities. In contrast, the mortality rate ($ \mu $)
has a rather narrow range of variation. For this reason,
if there is to be a relationship between $ \mu $ and $ d $
one would expect $ \mu $ to depend upon $ \log d $. This is the point
already emphasized in the introduction when we said that one
needs to consider large changes of $ d $. Does this
suffice to reveal a definite correlation?
Fig. 1c and Table 2 show that this is indeed  true
at the level of US states for several
contagious diseases; yet influenza stands as an exception
as shown by Fig. 1b.

%
\begin{table}[htb]

\small
\centerline{\bf Table 2: Impact of the population density $ d $  on 
the death rate $ \mu $ of contagious diseases, US states}

\vskip 5mm
\hrule
\vskip 0.7mm
\hrule
\vskip 0.5mm
$$ \matrix{
&\hbox{}\hfill & \hbox{Coefficient}\hfill & \hbox{Exponent} \cr
&\hbox{}\hfill & \hbox{of} & \hbox{of the} \cr
&\hbox{}\hfill & \hbox{correlation}\hfill & \hbox{power law} \cr
\qtb
&\hbox{}\hfill & \hbox{}\hfill & \mu=Cd^{\alpha} \cr
\noalign{\hrule}
\qth
1&\hbox{Measles, 1915}\hfill & 0.71& 0.35 \pm 0.17\cr
&\hbox{Measles, 1918}\hfill & 0.47 & 0.20\pm 0.14\cr
&\hbox{\color{red} Measles, average}\hfill & & \color{red} 0.28\pm 0.11\cr
2&\hbox{Diphtheria 1915}\hfill & 0.67 & 0.24\pm0.11\cr
&\hbox{Diphtheria 1918}\hfill & 0.56 & 0.19\pm0.10\cr
&\hbox{\color{red} Diphtheria, average}\hfill & & \color{red} 0.22\pm 0.07\cr
3&\hbox{Whooping cough, 1915}\hfill & 0.13 &0.04\pm 0.12 \cr
&\hbox{Whooping cough, 1915}\hfill & 0.41& 0.17\pm0.14\cr
&\hbox{\color{red} Whooping cough, average}\hfill & & \color{red}0.11\pm 0.09\cr
4&\hbox{Pneumonia, 1915}\hfill & 0.59&0.10\pm 0.06 \cr
&\hbox{Pneumonia, 1918}\hfill & 0.60& 0.17\pm 0.08\cr
&\hbox{\color{red} Pneumonia, average}\hfill & & \color{red}0.14\pm 0.05\cr
5&\hbox{Tuberculosis, 1915}\hfill & 0.39&0.12\pm 0.11 \cr
&\hbox{Tuberculosis, 1918}\hfill & 0.48&0.15\pm 0.10 \cr
\qtb
&\hbox{\color{red}Tuberculosis, average}\hfill & & \color{red}0.14\pm 0.07 \cr
\noalign{\hrule}
} $$
\vskip 1.5mm
Notes: The correlations and regressions are for $ (\log d,\log \mu) $.
Taking the log of $ \mu $ is not a necessity (for $ \mu $ has
a small range of variation) but has the advantage of making the 
regression independent of the way $ \mu $ is measured (for
example per 1,000 or 100,000).
These estimates are based on the data of 
US registration states; there were 25 in 1915 and 30 in 1918. 
At this level there is no significant correlation
for influenza alone; however,
most often influenza and pneumonia
are counted together. 
Apart from 1918, in all ``normal'' years there were
about 10 times more pneumonia deaths than influenza deaths.
In 1918 the two diseases had about the same death rate.
Note that almost all these exponents are under 0.25
which suggest a fairly weak connection ($ \alpha=0 $
would mean no connection at all).
\qL
Source: Mortality statistics 1919; this volume has
a recapitulation for the years 1915 to 1919.
\vskip 2mm
\hrule
\vskip 0.7mm
\hrule
\end{table}

The results given in Table 2 show that, at least in this
time period, the values of the
exponent of the power law were fairly stable in the course
of time. The exponent found in the next subsection for
the influenza epidemic of 1918, namely $ 0.22 $ is in the same
range.
\qpar

It must be emphasized that exponents $ \alpha $
in the range $ 0.10-0.25 $
denote a fairly weak interdependence (obviously for
$ \alpha=0 $ there would be no relationship at all).
That is why this effect can be easily covered by the
background noise.

\qA{Influenza-pneumonia epidemic}

Thanks to a special report published by the US Bureau
of the Census (1920) which describes the spread of the
influenza epidemic in the fall of 1918 
we have far more detailed data in this case than for any
other. As in addition this epidemic was particularly strong
the relative magnitude of the background noise will
be reduced thus providing excellent observation conditions.
\qpar

Fig. 2 summarizes the situation. Whereas there is a marked
density-death rate correlation ($ r=0.90 $) on a broad
density scale, within rural or urban places it is the
background noise which dominates.

%
\begin{figure}[htb]
\centerline{\psfig{width=11cm,figure=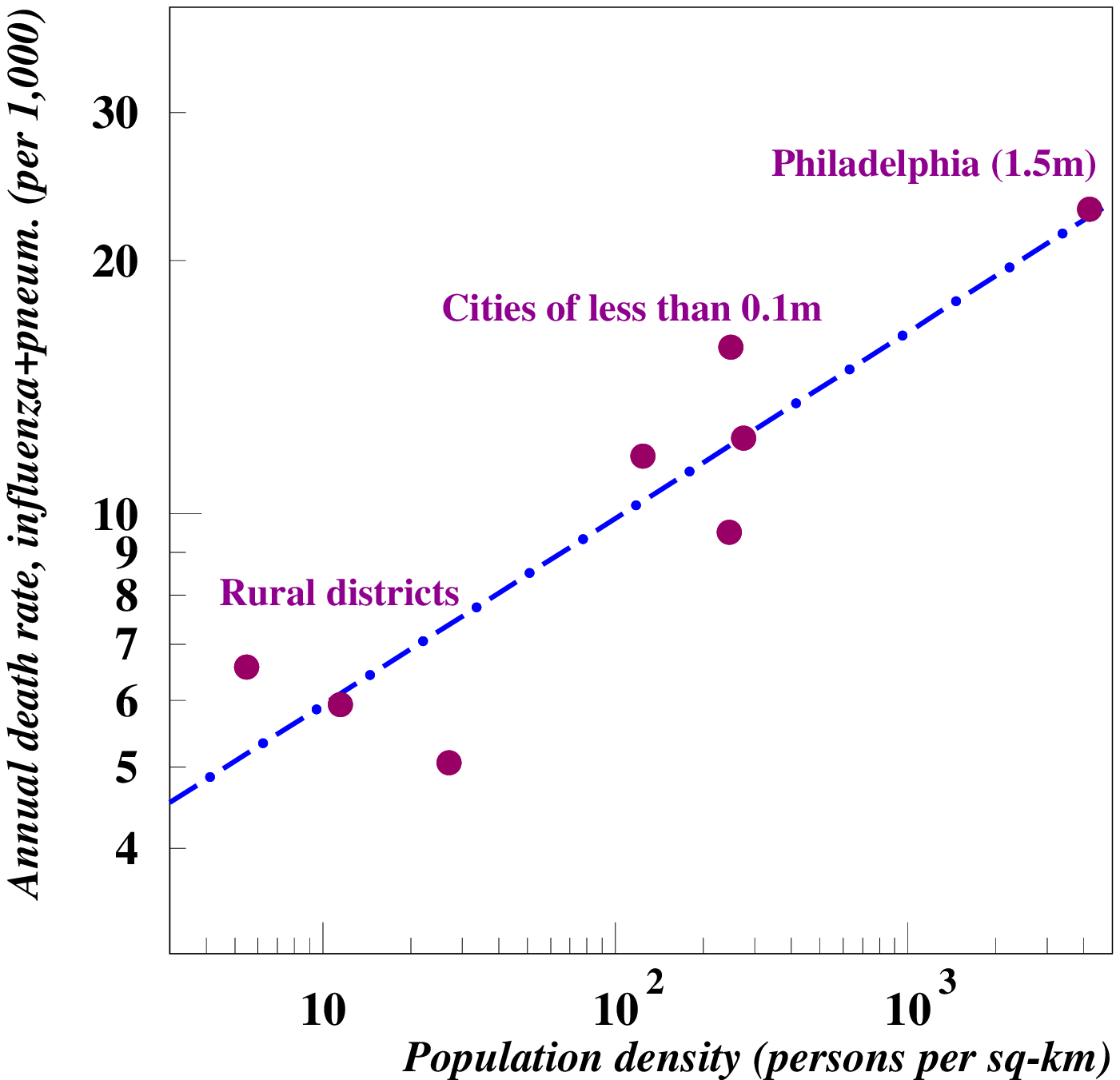}}
\qleg{Fig.\qhu 2\qhv Relationship between population density $ d $
and the size $ \mu $ of the influenza epidemic of September-December 1918.}
{In the graph $ m $ means million.
The data are for Indiana, Kansas and the city of Philadelphia
in Pennsylvania. Influenza and pneumonia deaths are counted together.
It can be seen that the relationship
between population density holds only on a broad density scale.
Inside of the three groups of data points the background fluctuations
are strong enough to override the power law. The regression
reads (the confidence interval is for a confidence probability 
of 0.95):
$ \mu=Cd^{\alpha},\ \alpha=0.22\pm0.08,\ C=3.5 $.}
{Source: Bureau of the Census (1920).}
\end{figure}
%
\qA{Effect of population density on the evolution of the 
epidemic}

Fig. 3 shows that the shape of the evolution curves
is very much density dependent. A high density gives a sharp
peak whereas low density gives broad bulges.  
%
\begin{figure}[htb]
\centerline{\psfig{width=11cm,figure=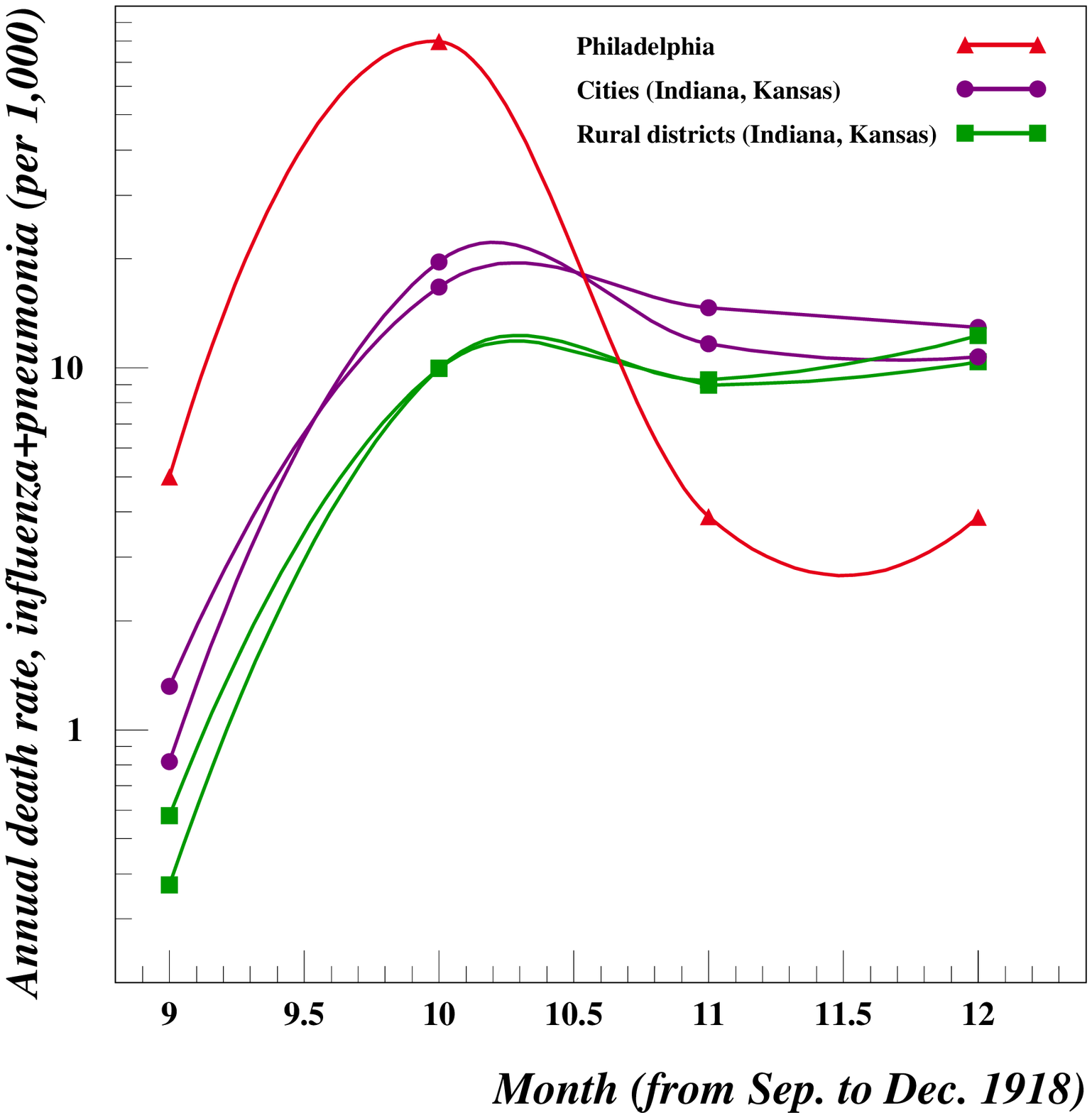}}
\qleg{Fig.\qhu 3\qhv Evolution of the death rate of
the influenza epidemic from September to December 1918.}
{It is remarkable that the curves for Indiana and
Kansas are almost the same in spite of a distance of
about 1,000 km between them; in contrast the curves
are very dependent upon the population density.}
{Source: Bureau of the Census (1920).}
\end{figure}
%

\qI{Determinants of an epidemic: the KMK model}

What is our purpose in discussing the KMK model
in relation with the question of the impact of population
density? 
\qpar

At the beginning of the paper we 
explained that the it is the question of the contacts
between native populations and immigrants which motivated
the present study. However, this question cannot be settled
in a definitive way by relying only on empirical evidence.
The reason is that together with their various strains
contagious diseases form an unclosed set. If one adds
to this uncertainty that one has almost no information
about the immune systems of native populations
it becomes obvious that investigation of specific case-studies
can hardly give us an overall understanding.
\qpar
 
It is by pupose that the model presented in this section
involves only the most basic features of an epidemic,
namely contagion, recovery and death. In this way our
hope is to capture and understand the very mechanism of
epidemics. The fact that local conditions usually do not
play a great role is demonstrated by the similarity of
the course of the influenza epidemic (one of the
few for which extensive daily data are available)  
in various cities whether in Europe or in the United States.

\qA{Parameters and differential equations}

A simplified model of an epidemic can be seen as defined by
3 parameters (see Fig. 4)
\qee{1} An infection (or incidence) rate, $ \beta $, which describes
the transition from health to illness.
\qee{2} A removal rate, $ \gamma $, which describes the transition
from illness to death or recovery.
\qee{3} A fatality rate, $ \gamma_1 $, which defines the
proportion of deaths in the wake of the disease.
\qpar

In the argument which leads to the equations defining the model
the crux of the
matter is the fact that newly infected persons are
generated through interaction between a person that is
already infected (described by the variable $ y $)
and a susceptible person, i.e., a person
not yet infected and who has not yet developed an
immunity (described by the variable $ x $). 
In the differential equation of
the model this interaction is described by a product term 
$ \beta xy $ where $ \beta $ describes the infection process.
As the disease progresses the pool of infected persons
is depleted because infected persons may die or may
recover and then be immune at least for the near future.
This removal process will be described by a term $ -\gamma y $.
In other words there is a competition between infection
and removal which can be quantified by the ratio 
$ \rho=x_0\beta /\gamma $. 

%
\begin{figure}[htb]
\centerline{\psfig{width=8cm,figure=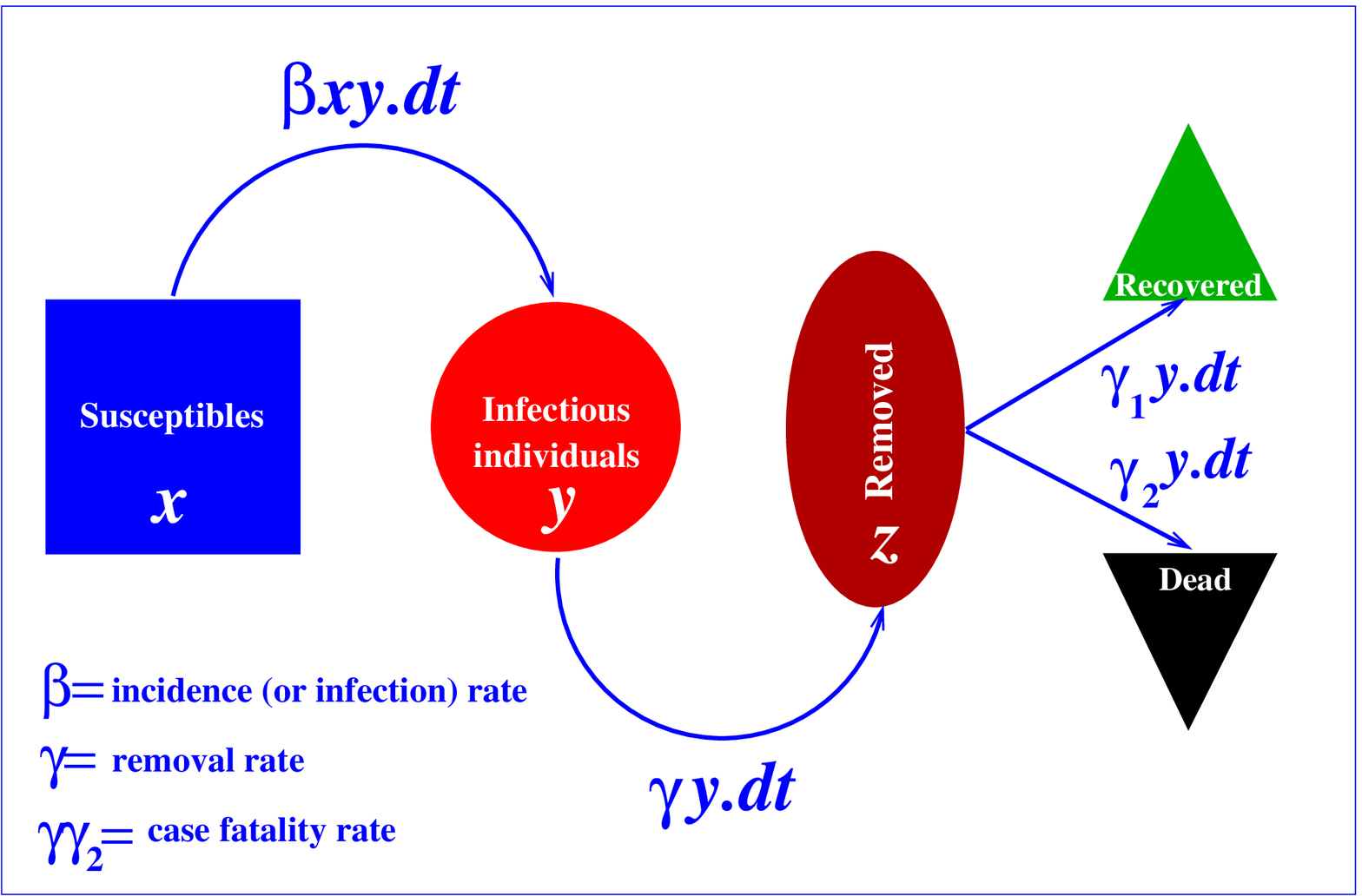}}
\qleg{Fig.\qhu 4\qhv Diagram illustrating the mechanism of the
KMK model (Kermack and McKendrick 1927)
for the propagation of an epidemic.}
{$ x(t)= $persons susceptible to infection, $ y(t)= $infected (and
  infectious) persons, $ z(t)= $persons who have been infected
and whi, at time $ t $ are either dead or immune to infection.}
{}
\end{figure}
%

This is summarized in the following 
system of differential equations.
$$ \left\{\matrix{
dx/dt & = & -\beta xy & & &\quad & (1.1)\cr
dy/dt & = & \hfill \beta xy & - & \gamma y & \quad & (1.2)\cr
dz/dt & = & & & \gamma y & \quad & (1.3) \cr
}
\right. \leqno(1) $$

In addition it should be added that $ x+y+z=n $ and 
that we are only interested
in non-negative solutions, that is to say: $ x(t),y(t),z(t)\ge 0 $.
\qpar

It can easily be shown that $ z $ satisfies the following non linear
equation:
$$ { dz\over dt }=\gamma \left[n-
x_0\exp\left(-{ \beta \over \gamma }z \right) -z \right] \qn{2} $$

By expanding the exponential to second order one gets a logistic
equation which can be solved analytically. This quadratic
approximation is valid when $ \beta/\gamma \ll 1 $  and remains
valid as long as $ z $ is small enough. 
\qpar

The infection and removal effects are very different from one 
another
\qbu $ \beta $ is determined by the type of contagion which
is a biological factor but it 
is also highly dependent upon the frequency of inter-individual 
contacts%
\qfoot{The relationship between the network structure
of the population and the frequency of interactions was
examined in Li et al. (2013).}%
. 
It can be expected to be small when the population density is
low. 
\qbu $ \gamma $ describes the evolution of the disease
either to death or to recovery. Thus, it is chiefly a biological
parameter which is dependent upon the type of the disease.

\qA{Size of the epidemic}

When $ t \rightarrow \infty $ the variable $ z $ which represents
the persons affected by the epidemic converges toward a stationary 
limit which is the solution of the equation:
 $$ n-z=x_0\exp\left[-(\beta/\gamma)z \right] \qn{3} $$

%
\begin{figure}[htb]
\centerline{\psfig{width=17cm,figure=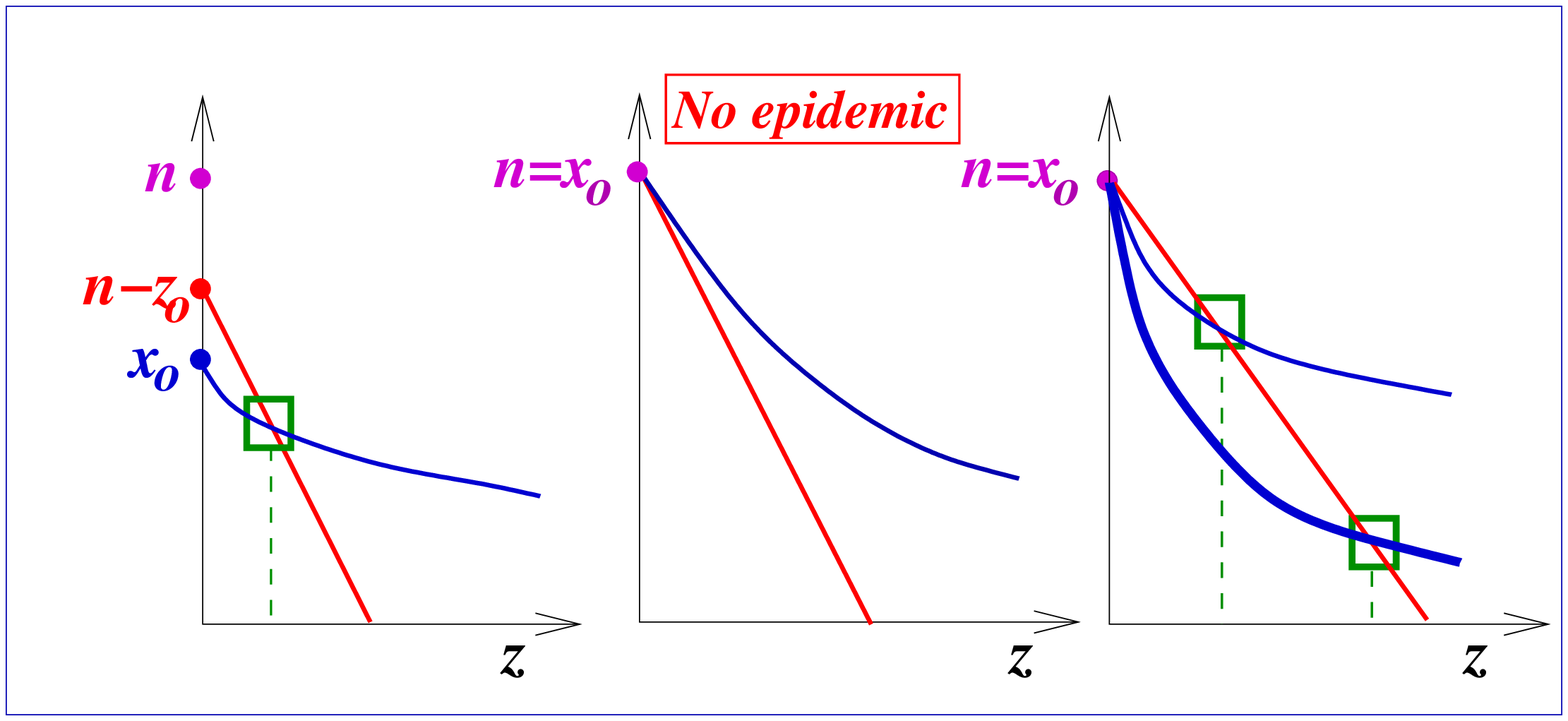}}
\qleg{Fig.\qhu 5a,b,c \qhv Limiting value of the variable
z.}
{The straight line represents the left-hand side of
equation (3) whereas the curve represents the exponential
in the right-hand side of the same equation. The intersections
marked by the green squares correspond to the asymptotic 
value $ z(t\rightarrow\infty) $. {\bf (a)} This figure
shows that when $ y_0+z_0>0 $ there is always an intersection.
{\bf (b)} On the contrary if $ y_0+z_0=0 $, then the
epidemic cannot start. {\bf (c)} This figure shows
that the intersection moves to the right when the threshold
parameter increases.}
{}
\end{figure}

The functions on the left- and right-hand size of this equation
are shown in Fig. 5a,b,c. Fig. 5a shows that when $ y_0+z_0>0 $ it
has always a solution whatever the value of $ \rho $;
however when $ \rho $ is close to 1 the epidemic may be small.
Fig. 5b shows that when $ y_0+z_0=0 $, then there is no
epidemic unless $ \rho $ is larger than 1. Finally, Fig. 5c
shows that the limit of $ z $ increases when $ \rho $ becomes
larger. The way the size of the epidemic increases with $ \rho $
is shown more precisely in Fig. 6b which is based on a 
numerical solution of equation (2). 
\qpar

With respect to the question of whether a population can
be wiped out by an epidemic which will be discussed in the
conclusion it is important to observe that even if none of
the infected persons recovers (which corresponds to $ \gamma_1=0 $
in Fig. 4) the population will {\it not} be wiped out.
Only a fraction of it will die, although it is true that
this fraction
may become close to one when $ \rho $ becomes 
very large. 

\qA{Key-role of population density}

The threshold parameter $ \rho=x_0\beta/\gamma $ is
proportional to the initial number of susceptibles 
which itself, in case of a new disease, is close to the total  
population. The model does not describe the spatial
aspects of the epidemic but as it is formulated
for a population $ n $ on 
a given territory it implies that population
density and total population are proportional.
In other words, $ n $ plays the role of population density.
\qpar

Fig. 6b shows how the size of the epidemic increases
with the threshold parameter that is to say, 
whenever $ x_0\simeq n $, with population density. \qL
In the real world,
one expects $ \beta $ also to increase with population
density. As explained earlier, $ \beta $ depends 
upon the number of contacts and one expects people
to have more interactions (in stores, public transportation,
entertainment places
or at work) in cities than in rural places. Needless
to say, the level of $ \beta $ in cities depends upon
the special features of the city%
\qfoot{The hydrological environment plays a major role
in the spreading of cholera. More generally,
the role of population
distribution and of human interaction
intensity was examined in Li et al. (2017 a,b).}%
. 
For instance,
because of the difference in public transportation
one would expect $ \beta $ to be higher in Tokyo
than in Los Angeles.

%
\begin{figure}[htb]
\centerline{\psfig{width=17cm,figure=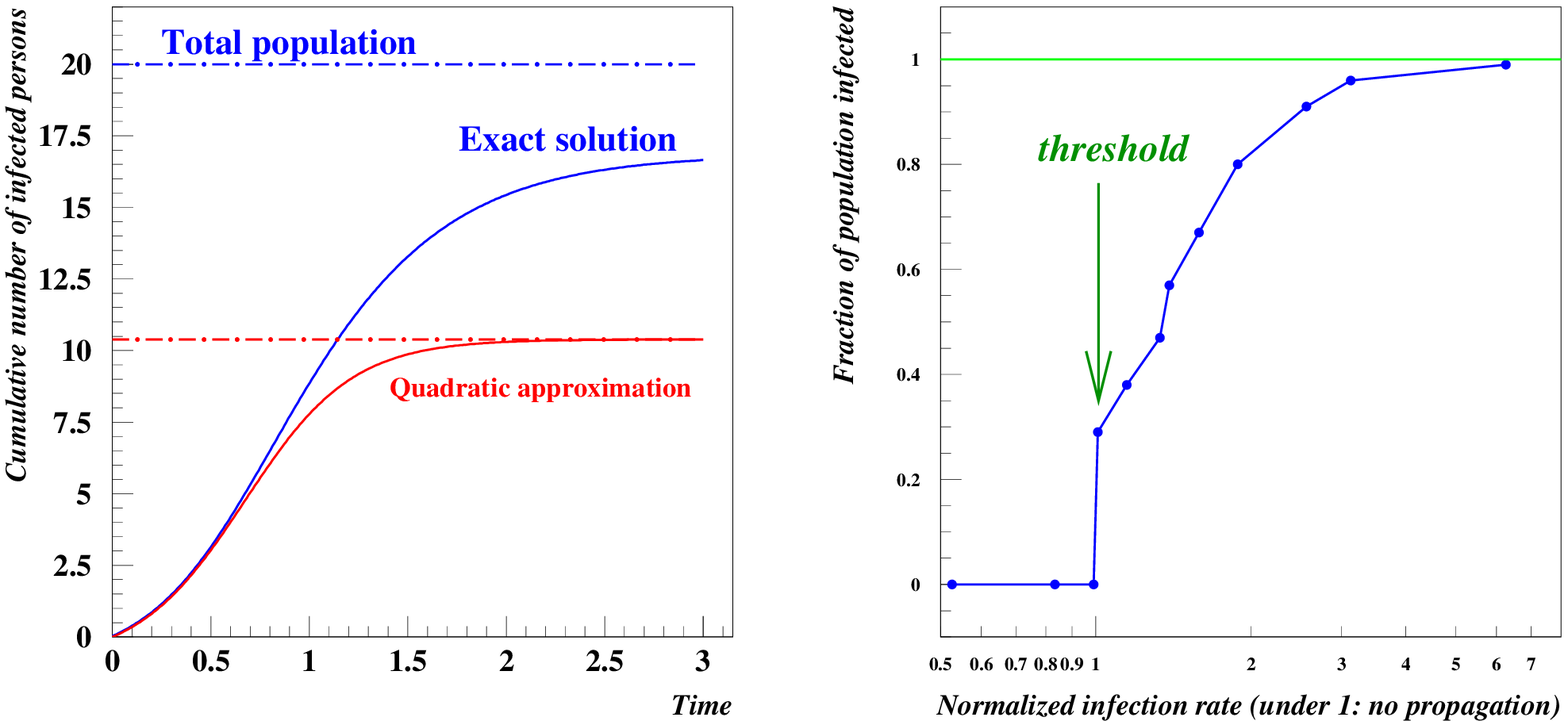}}
\qleg{Fig.\qhu 6a,b \qhv The KMK model.}
{{\bf (a)} Increase in time of the fraction of the 
population which has been in contact with the disease.
This simulation corresponds to the following
parameters: total population: $ n=20 $, $ x_0=n-1,\ \beta=0.32,\
\gamma=3,\ \rho=x_0\beta/\gamma=2.03 $. The model's equations
must be solved numerically, but there is also an analytic
approximation which is shown by the lower curve.
The accuracy of this approximation is controlled
by the threshold parameter $ \rho $. When $ \rho $ is slightly
larger than 1, the infection starts slowly and only a small
fraction of the population becomes infected. 
{\bf (b)} This graph shows the total fraction of the population
that has become infected as a function of the threshold
parameter. The infection can start as soon as $ \rho> 1 $;
when $ \rho > 2.5 $ the fraction infected is over 90\%.}
{}
\end{figure}
 
{\bf Remark}\qL
It can be added that the increase of the size of the epidemic
with the density is specific to the exact model.
In the quadratic approximation (which results in a logistic
equation for $ z(t) $) the size of the epidemic (that is to say
the limit of $ z(t) $ when $ t $ goes to infinity) is given
by the expression: 
$$ z(\infty)=2\gamma /(x_0\beta)\left[x_0-\gamma/\beta \right] $$
which, obviously, does not increase with $ x_0 $.

\qA{Comparison to observation}
%
\begin{figure}[htb]
\centerline{\psfig{width=12cm,figure=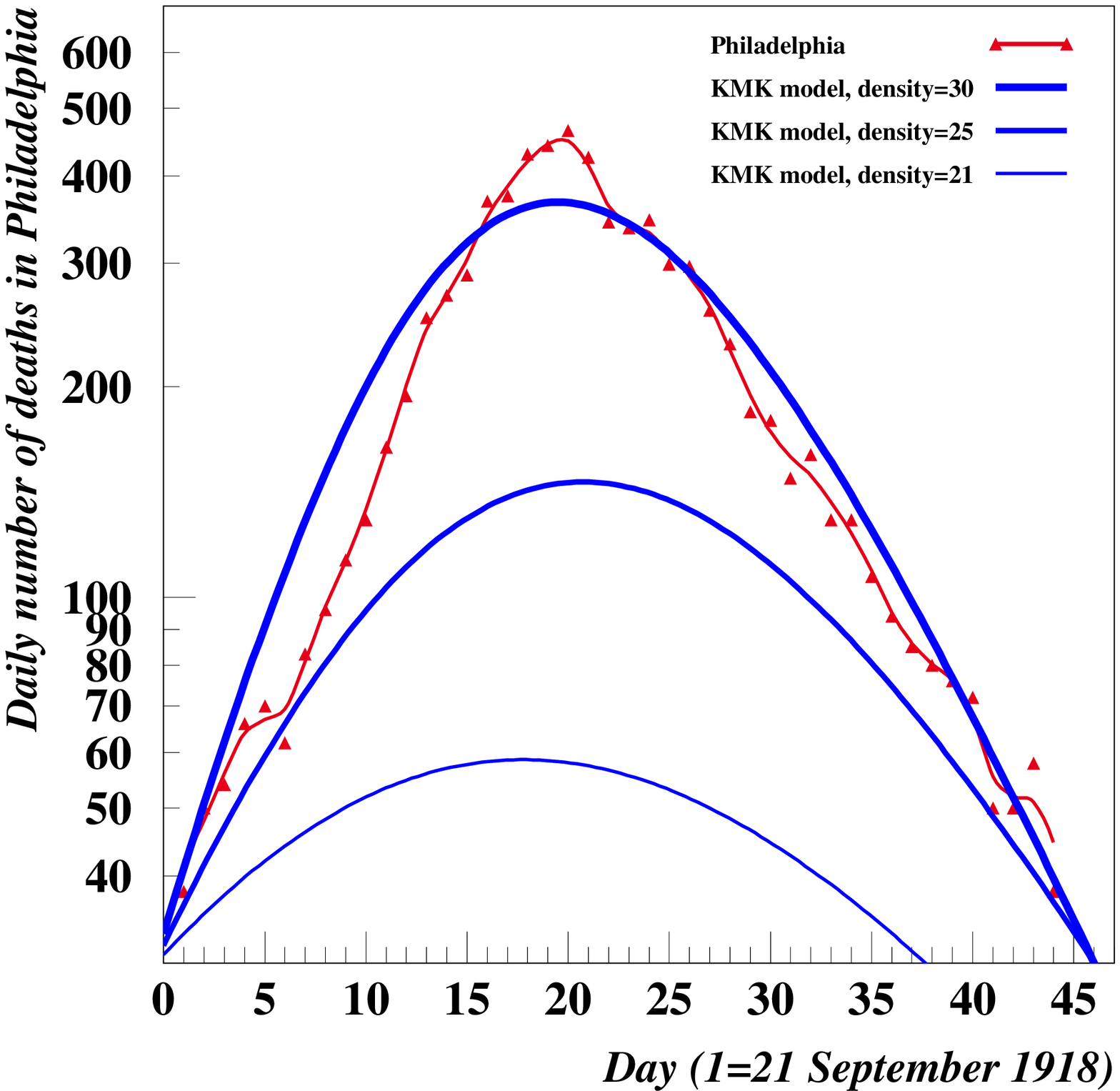}}
\qleg{Fig.\qhu 7\qhv Predictions of the KMK equations
for different densities and comparison with 
daily deaths in Philadelphia.}
{It can be seen that for the model as well as for
Philadelphia the raising and falling parts are
nearly exponential.
Apart from the population $ n $, the other parameters 
have the following values: 
$ \beta=0.32,\ \gamma=5,\ x_0=n-1,\ y_0=1,\ z_0=0 $.}
{Source: The daily data for Philadelphia are from: Bureau
of the Census (1920).}
\end{figure}
%
The ``Special report'' gives death rates by month.
In a few instances it gives
also daily death data which provide a more accurate
view of the shape of the curve. In Fig. 7 we tried to 
determine parameters which would lead to this shape.
The height of the peak can be easily controlled through the
threshold parameter; this is shown in fig. 7 by the
three curves corresponding to different densities.
However, a larger $ \rho $ will
give a curve whose falling part is wider than its raising
part whereas in fact the empirical curve is almost 
symmetrical with respect to its peak value. 
The descending part can be made shorter by increasing
$ \gamma $. In this way, we can define a set of parameters which
approximates fairly closely the empirical curve. 

\qA{Predictions of the model}

If we really believe in this model we should use it
to make testable predictions. One of its main features
is the existence of a threshold under which the death
rate falls very abruptly. In other words for sufficiently
low densities one should see a sudden drop of the death 
rates. Practically, however, what we can see is limited
by the noise. In our comments about Fig. 2 we have already
observed that for rural places the impact of the density
becomes over-ridden by the noise. Note that the problem
of the noise is more serious for low densities than
for high densities because low densities means 
few deaths which in turn imply high statistical fluctuations.
As observed in the first section, such fluctuations
come in addition to the background noise.\qL
Thus, there is little hope that this threshold transition
will be observable.

\qI{Conclusion}

In this paper it was shown that there is a weak but 
clearly defined relationship
between population density and the death rate of epidemics
provided that sufficiently large density changes are
considered and background noise is kept under control.
It was also shown that
population density determines
the time dependence of the death rate; thus, large densities
(as in Philadelphia) lead to high narrow peaks
whereas for small densities one observes
low and broad humps.
\qpar

The question of the length of time of an epidemic
process would deserve a closer study. In the present
paper we considered only the case of influenza + pneumonia.
which are diseases for which the incubation time and the length of
survival may be as short as a few days. However, for
other diseases these times may be much longer: for rabies
it is a few months, for AIDS a few years%
\qfoot{Rabies and AIDS have specific spreading mechanisms
which should be taken into account in any model description.}%
Needless to say, the longer the process, the more difficult
it is to determine its length.  
\qpar

In a recent paper (Richmond et al. 2018) a methodology was
developed which allows measurement of the strength of 
family interactions
between spouses or between parents and children.
One may wonder whether the propagation of a disease
can serve to estimate the proximity between family
members and more broadly between people. At this point the only
thing one can say is that this would require
detailed epidemic micro data that do not seem available.
\qpar
%

Epidemics ascribed to a lack of immunity in native
populations are often given as the reason of their collapse.
The following excerpt taken from Marsh (2004) is typical
of this kind of statements:
\qdec{``Nevada Indians had no immunity to the diseases that white explorers,
colonists and settlers brought to their lands. These diseases included
smallpox, measles, tuberculosis and others, which ravaged the tribes
in great epidemics that killed many, and sometimes all, members of a
tribe''.}
\qpar

From a scientific point of view such statements are unsatisfactory
for at least three reasons.
\qee{1} It is not easy to determine the moment when 
a native population has come
in contact with persons who may carry pathogens. For instance,
it is clear that the Nevada Indians have had contacts with
Spanish people for a long time before the area became part of the
US following the Mexican-American War of 1846-1848. The main
difficulty is that the paucity of sources  does not allow
us to set a date in a reliable way%
\qfoot{Actually, the very definition of the notion of  ``contact'' 
is unclear.
Is the arrival of one or several hunters sufficient to
start an epidemic? We do not know.
However, one can be sure that irrespective of the
initial contact, the disease will spread fairly slowly in low
density areas like Alaska, Arizona or Nevada.}
\qee{2} Most often native populations have low density.
This is of course true for the Nevada Indians. If one takes
$ d=1 $ person per sq.km as a rough density estimate of
native populations and $ d=100 $ as the density of present-day
France, the death rate due to a contagious disease will
be $ 100^{0.20}=2.5 $ times smaller in the native population.
From Table 2 we know
that the exponent is slightly disease-dependent; the value of
$ 0.20 $ taken here represents a rough average.
\qpar

In short, these two
features call into question the notion of sudden
collapses due to epidemics. 
\qee{3} Most often for native populations
 there are no census records nor reliable estimates.
However, in a few cases there are acceptable data going
back to the early 19th century; Alaska and the Tonga Islands
in the Pacific are two such cases and, remarkably,
their population did not experience any collapse after
coming into contact with white travelers. Below
we give some additional details for Alaska.
\qee{4} There are indeed documented cases of
sudden collapses within one or two decades. 
If diseases are not the right
explanation how can we explain them? \qL
There are plenty of possible reasons:
starvation when the traditional source of food
(e.g., vegetables, salmons, buffaloes) is no
longer available, dispersion of tribes and splitting
of families which prevents conceptions, or outright
killings. Such events can occur simultaneously
as documented by Benjamin Madley (2008, 2016)
for the California Indians.
\qpar

For the case of the Alaskan Indians there are two
conflicting accounts: Mooney (1929) sees a sharp population
fall due to diseases
over the period 1740-1780, a time interval for which there are
in fact no data available
whereas Petroff (1884) bases his account on the
population estimates which became available after 1780;
these do not show any sizable population decrease
in spite of the fact that the tribes of continental
Alaska came into contact with white people only by 1840.
In other words, in this case there was no immunity shock.

\vskip 5mm
{\bf References}

\qparr
Bureau of the Census 1920: Special tables of mortality from influenza
and pneumonia. Indiana, Kansas, and Philadelphia, September 1 to
December 31, 1918. Government Printing Office, Washington DC.

\qparr
Chowell (G.), Bettencourt (L.M.A.), Johnson (N.), Alonso (W.J.), 
Viboud (C.) 2008:
The 1918–-1919 influenza pandemic in England and Wales: spatial
patterns in transmissibility and mortality impact.
Proceedings of the Royal Society B, 7 March 2008.

\qparr
Haidari (A.A.), Hazmi (A.A.), Marzouq (H.A.), Armstrong (M.),
Colman (A.), Fancourt (N.), McSweeny (K.), Naidoo (M.), Nelson (P.),
Parnell (M.), Rangihuna-Winekerei (D.), Selvarajah (Y.), 
Stantiall (S.), Wilson (N.), Baker (M.) 2006: 
Death by numbers. New Zealand mortality rates in the 1918
influenza pandemic. Publication of the Wellington School of
Medicine and Health Sciences. 

\qparr
Kermack (W.O.), McKendrick (A.G.) 1927:
Contributions to the mathematical theory of epidemics. 
Proceedings of the Royal Society, A, 115,700-721.

\qparr
Li (R.), Tang (M.), Hui (P.-K.) 2013: Epidemic spreading
on multi-relational networks.
Acta Physica Sinica, 62,16,168903.

\qparr
Li (R.), Wang (W.), Di (Z.) 2017a: Effects on human dynamics
on epidemic spreading in C\^ote d'Ivoire.
Physica A, 467,30-40.

\qparr
Li (R.), Dong (L.), Zhang (J.), Wang (X.), Wang (W.-X), Di (Z.),
Stanley (H.E.) 2017b: Simple spatial scaling rules behind
complex cities. Nature Communications, 8,1841.

\qparr
Linder (F.E.), Grove (R.D.) 1947: Vital statistics rates in the
United States, 1900-1940. National Office of Vital Statistics,
Government Printing Office, Washington DC.

\qparr
Madley (B.) 2008: 
California's Yuki Indians. Defining genocide in native American history
Western Historical Quarterly 39,3,303–332.

\qparr
Madley (B.) 2016: 
An American Genocide.
The United States and the California Indian Catastrophe, 1846-1873.
Yale University Press, New Haven.

\qparr
Marsh (C.) 2004: Nevada native Americans. Galopade International.

\qparr
Mooney (J.) 1929:  The aboriginal population of America north of Mexico.
Smithsonian Miscellaneous Collections Vol.80, No 7,1-40. 

\qparr
Petroff (I.) 1884: Report on the population, industries, and resources
of Alaska. Prepared at the request of the US Bureau of the Census.

\qparr
Richmond (P.), Roehner (B.M.) 2018: Exploration of the
strength of family links.
Physica A 502,1-13.

\end{document}